\documentclass[11pt,a4paper]{article}
\usepackage{jheppub}

\usepackage{graphicx}
\usepackage{dcolumn}
\usepackage{bm}
\usepackage{color}

\begin{document}

\title{Single electron emission in two-phase xenon \\
with application to the detection of coherent \\
neutrino-nucleus scattering}

\collaboration{ZEPLIN--III Collaboration}

\author[1,2]{E.~Santos,}
\author[3]{B.~Edwards,}
\author[1]{V.~Chepel,}
\author[2]{H.M.~Ara\'{u}jo,}
\author[4]{D.Yu.~Akimov,}
\author[5]{E.J.~Barnes,}
\author[4]{V.A.~Belov,}
\author[4]{A.A.~Burenkov,}
\author[2]{A.~Currie,}
\author[1]{L.~DeViveiros,}
\author[5]{C.~Ghag,}
\author[5]{A.~Hollingsworth,}
\author[2]{M.~Horn,}
\author[3]{G.E.~Kalmus,}
\author[4]{A.S.~Kobyakin,}
\author[4]{A.G.~Kovalenko,}
\author[2]{V.N.~Lebedenko,}
\author[1,3]{A.~Lindote,}
\author[1]{M.I.~Lopes,}
\author[3]{R.~L\"{u}scher,}
\author[3]{P.~Majewski,}
\author[5]{A.St\,J.~Murphy,}
\author[1,2]{F.~Neves,}
\author[3]{S.M.~Paling,}
\author[1]{J.~Pinto da Cunha,}
\author[3]{R.~Preece,}
\author[2]{J.J.~Quenby,}
\author[5]{L.~Reichhart,}
\author[5]{P.R.~Scovell,}
\author[1]{C.~Silva,}
\author[1]{V.N.~Solovov,}
\author[3]{N.J.T.~Smith,}
\author[3]{P.F.~Smith,}
\author[4]{V.N.~Stekhanov,}
\author[2]{T.J.~Sumner,}
\author[2]{C.~Thorne}
\author[2]{\& R.J.~Walker}

\affiliation[1]{LIP--Coimbra \& Department of Physics of the University of Coimbra, Portugal}
\affiliation[2]{High Energy Physics group, Blackett Laboratory, Imperial College London, UK}
\affiliation[3]{Particle Physics Department, STFC Rutherford Appleton Laboratory, Chilton, UK}
\affiliation[4]{Institute for Theoretical and Experimental Physics, Moscow, Russia}
\affiliation[5]{School of Physics \& Astronomy, University of Edinburgh, UK}

\emailAdd{e.santos10@imperial.ac.uk}

\abstract{We present an experimental study of single electron emission
in ZEPLIN--III, a two-phase xenon experiment built to search for dark
matter WIMPs, and discuss applications enabled by the excellent
signal-to-noise ratio achieved in detecting this signature. Firstly,
we demonstrate a practical method for precise measurement of the free
electron lifetime in liquid xenon during normal operation of these
detectors. Then, using a realistic detector response model and
backgrounds, we assess the feasibility of deploying such an instrument
for measuring coherent neutrino-nucleus elastic scattering using the
ionisation channel in the few-electron regime. We conclude that it
should be possible to measure this elusive neutrino signature above an
ionisation threshold of $\sim$3 electrons both at a stopped pion
source and at a nuclear reactor. Detectable signal rates are larger in
the reactor case, but the triggered measurement and harder recoil
energy spectrum afforded by the accelerator source enable lower
overall background and fiducialisation of the active volume.}

\date{\today}


\notoc

\maketitle

\section{Introduction}

ZEPLIN--III is a two-phase xenon detector designed for the direct
detection of WIMP dark matter \cite{sumner01,akimov07}. The two-phase
technique \cite{dolgoshein70,barabash89,bolozdynya95} produces two
different signals for each particle interacting within the active
volume, one from primary scintillation in the liquid (S1) and the
other from electroluminescence in the gas phase (S2). This second
signal is a measure of the amount of ionisation drifted from the
interaction site by application of an electric field and subsequently
emitted from the liquid surface. The scintillation and ionisation
yields differ for electron and nuclear recoils, providing a physical
basis for the discrimination between interaction types. Since
discrimination requires both signals, in typical WIMP dark matter
searches the energy threshold is determined by the less sensitive S1
response channel.

The ionisation signal is measured through cross-phase emission of
electrons extracted from tracks in the liquid into the xenon vapour
above it. Here, they are accelerated by a strong electric field and
induce secondary scintillation (electroluminescence) in the xenon
vapour. The photon yield is a linear function of the field $E$ and can
be parametrised by $N_{ph}=(aE+bn)x$, where $x$ is the thickness of
the gas layer, $n$ is the number density of xenon atoms and the
coefficients $a$ and $b$ were determined experimentally for saturated
xenon vapour in Ref.~\cite{fonseca04}. For typical ZEPLIN--III
operational parameters ($E$=7--8~kV/cm in the gas at 1.6~bar,
$x\!\sim\!4$~mm), some 300 photons are produced by a single electron
emitted from the liquid.

The possibility of exploiting the sensitivity of the ionisation
channel down to the single electron level (sub-threshold in S1) has
been pointed out in the context of coherent neutrino-nucleus
scattering in 2004~\cite{hagmann04}. This motivated a programme based
on two-phase argon for nuclear reactor
monitoring~\cite{sangiorgio10}. Interest in this experimental
technique has now also extended to searches for light
WIMPs~\cite{angle11}. With this type of application in mind, we first
characterised the single electron signature using ZEPLIN--II
data~\cite{edwards08} and in work with a smaller prototype chamber
operating in a surface laboratory~\cite{burenkov09}.

This article is organised as follows. In Section~II we discuss the
single electron signature and likely production mechanisms based on
analyses of independent datasets acquired under different conditions.
This leads to the demonstration of a very practical application: the
determination of the free electron lifetime in the liquid phase by
using WIMP-search data only. This could be especially relevant for
next-generation WIMP experiments based on the noble liquids. In
Section~III we assess the feasibility of using the ionisation channel
down to single electron level to measure the coherent elastic
scattering of neutrinos off nuclei, a Standard Model process not yet
observed. We include realistic signal characteristics and a background
of single electron pulses which might be produced by the mechanisms
identified in the preceding section. Previous studies have typically
assumed idealised response models.

\section{Single electrons in ZEPLIN--III}

Single electron signals were studied using four independent datasets:
two WIMP-search exposures (of several months' duration) and two
dedicated runs. The main difference between the two types of dataset
is the origin of the trigger for data acquisition (DAQ). During dark
matter data-taking, the trigger function was derived from S1 or S2
pulses in the detector, recording $\pm$18~$\mu$s waveforms centred around
the trigger point. In the dedicated runs, on the other hand, the DAQ
was triggered externally with a pulse generator at a fixed (high)
repetition rate, independently of any signals in the detector. In this
instance the acquired waveforms were 256~$\mu$s long, which exceeds
significantly the maximum ionisation drift time in the chamber
($\sim$14.5~$\mu$s). We begin by describing the set-up and results
from the first dedicated single-electron run (hereafter `DSER'), which
underwent the most extensive analysis. Key parameters for this and
other datasets are summarised in Table~I.

\begin{figure}
\begin{center}
\includegraphics[height=5in,angle=270]{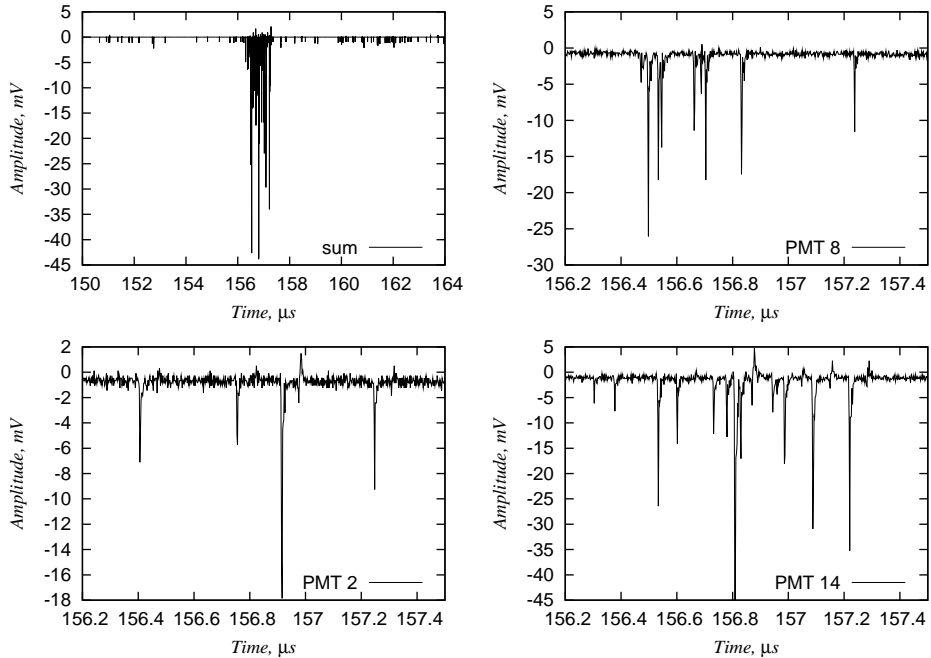}
      \caption{A single electron signal in ZEPLIN--III (sum waveform
      shown top left; channel waveforms containing detected
      photoelectrons are zoomed in).}
       \label{SEcluster}
\end{center}
\end{figure}

\subsection{Dedicated single electron run}
\label{DSER}

A total live time of 161.4~s was recorded during the DSER acquisition
over a 2-day period after the completion of the first science run
(FSR) in 2008. During that time, the xenon pressure in the detector
remained stable at 1.6~bar and $E$=7.6~kV/cm in the gas. The signals
from each of the 31 PMTs were digitised with 2~ns sampling and 8-bit
resolution over the 256~$\mu$s timelines. The DAQ system was forced to
trigger at the maximum rate for this configuration (18~s$^{-1}$) by an
external pulse generator. The raw waveform data were reduced with the
ZE3RA pulse analysis algorithms \cite{neves11}, using similar
parameters to those adopted for the FSR data, except that a Fourier
transform digital filter was applied to the waveforms to remove
coherent noise induced by the pulse generator. The reduced data were
then searched for photoelectron clusters, characteristic of single
electron emission, at a minimum of 3-fold coincidence in the PMTs,
with a pulse amplitude threshold of 4 times the rms noise in each
waveform. An example of such cluster is shown in
Figure~\ref{SEcluster}. In this particular analysis, the number of
single photoelectron pulses in a cluster is counted by the number of
threshold crossings individually in each channel using a constant
50-ns integration time. A channel-by-channel correction is then
applied to account for the dead-time effect thus introduced (a
constant-rate Poisson process is assumed).

\begin{table}
\begin{center}
  \caption{\small Single electron rates in ZEPLIN--III}
  \begin{tabular}{l|c|cc|c|c}
    \hline
    \hline
           & live time & drift field$^\dag$ & cath. wire & event rate$^\ddag$ & SE rate\\
    Run    & seconds   & kV/cm              & kV/cm      & s$^{-1}$           & s$^{-1}$    \\
    \hline
    DSER    & 161 & 3.8 & 60 & 6.8  & 5.7$^\ast$ \\
    FSR     & 194 & 3.8 & 60 & 6.8  & $\sim$20 \\
    SSR     & 10.2& 3.4 & 38 & 0.45 & $\sim$1 \\
    Cs-137  & 37  & 3.4 & 29 & 145  & 118$^\ast$
 \\
    \hline
    \hline
  \end{tabular}
  \begin{flushleft}
  \begin{small}
    $^\dag$ Electric field in the liquid xenon away from cathode.\\
    $^\ddag$ Higher energy background above normal trigger threshold (S2$>$4~electrons).\\
    $^\ast$ 20~$\mu$s inhibit period
  \end{small}
  \end{flushleft}
  \end{center}
\end{table}

The following selection cuts are applied in this analysis: i) no
pulses present in the preceding 20~$\mu$s in any waveform; this is
greater than the ionisation drift time for the full liquid depth and
ensures that selected events are not induced {\em directly} by a
previous energy deposition in the liquid xenon; ii) the reconstructed
$x,y$ position of the event must be within the central 60~mm radius of
xenon (1.3~kg active mass); this allows the use of a simple centroid
position reconstruction algorithm and ensures uniform light
collection.

\begin{figure}
  \begin{center}
  \includegraphics[width=2.8in]{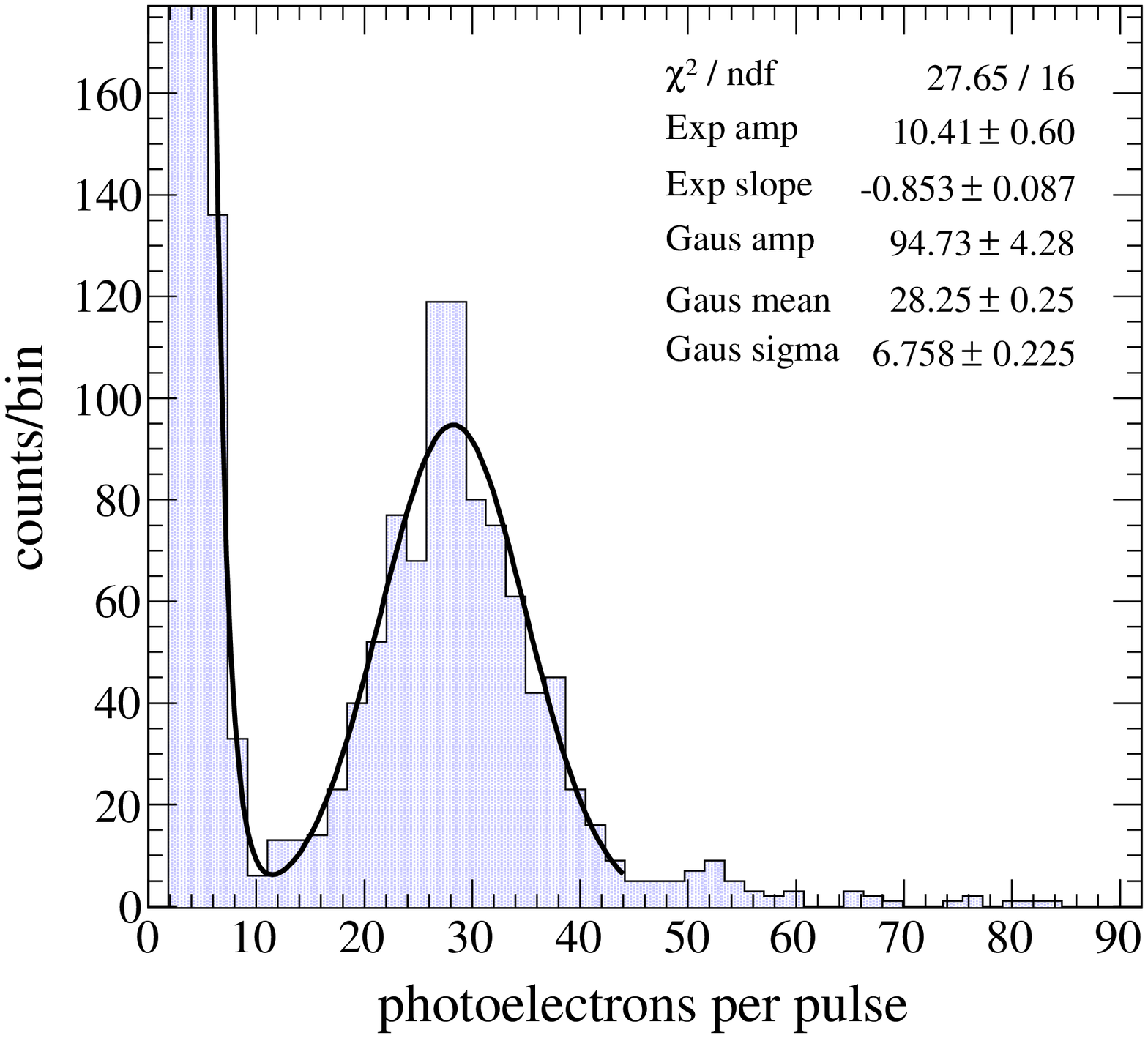}\quad
  \includegraphics[width=2.8in]{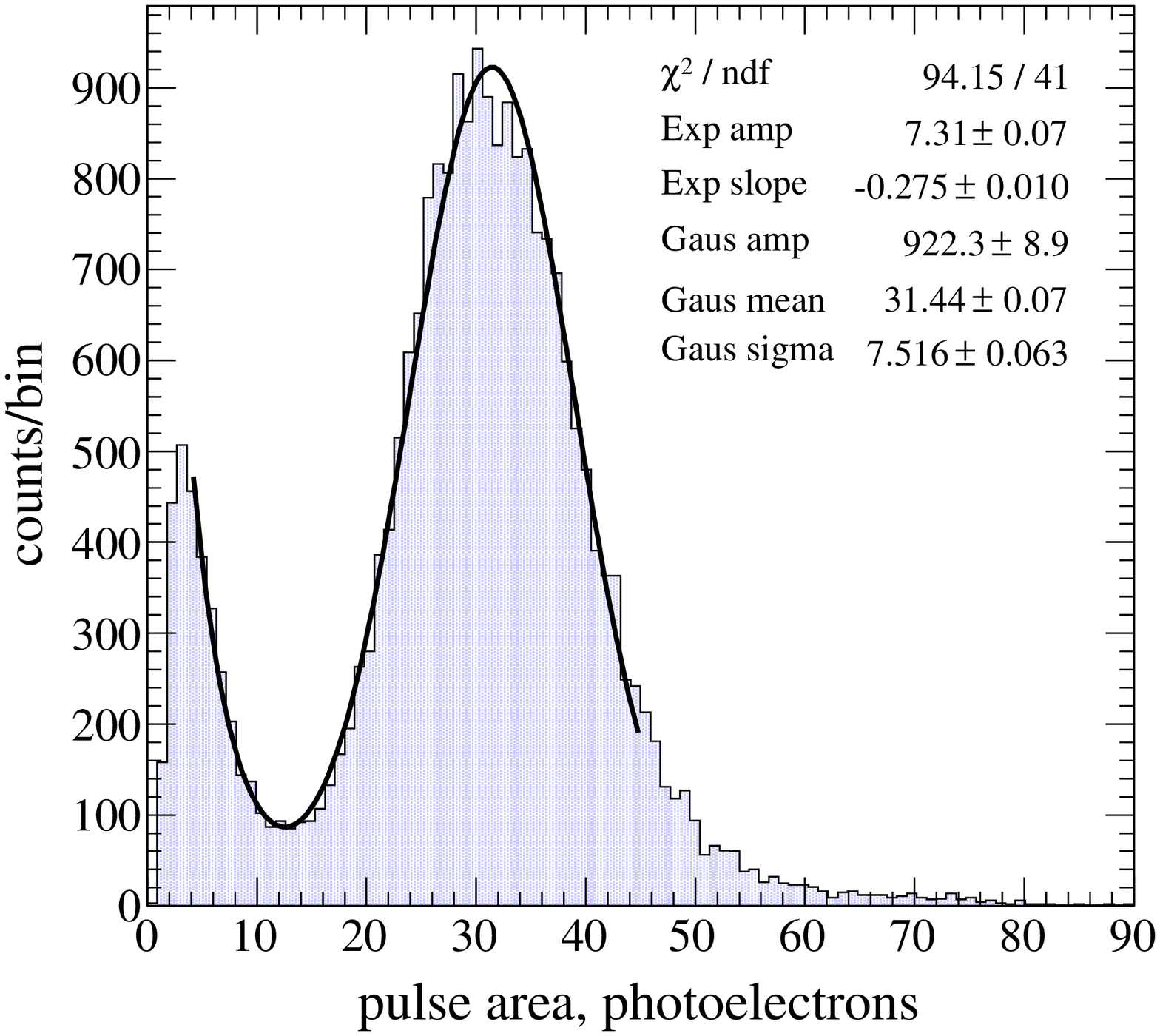}
  \end{center}
  \caption{Single electron pulse size distributions in the first
  science run configuration of ZEPLIN--III. \textsc{Left:} Number of
  photoelectrons per cluster obtained with the dedicated DSER dataset
  by photoelectron counting (dead-time corrected), as described in
  \S\ref{DSER}. \textsc{Right:} Size distribution of post-S1 pulses in
  the FSR science data, obtained by pulse area integration and
  converted subsequently into photoelectron numbers
  (\S\ref{postS1}).}
  \label{SEhistos}
\end{figure}

\begin{figure}
  \begin{center}
  \includegraphics[width=2.7in,clip=on]{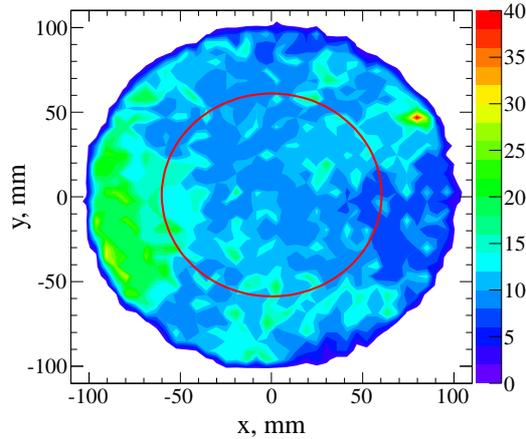}
  \end{center}
  \caption{Centroid-reconstructed position of `spontaneous' single
  electron pulses in the DSER dataset. This analysis extends to 60~mm
  radius. The bias observed at large radii is due to a small detector
  tilt which increases the gas layer thickness where an excess of
  events is observed.}
  \label{SEposition}
\end{figure}

Figure~\ref{SEhistos} (left) shows the size distribution of
photoelectron counts in single electron clusters for this dataset. The
histogram was fitted by exponential and Gaussian components, resulting
in a mean number of 28.3$\pm$0.3(stat) photoelectrons per pulse. The
very large peak-to-valley ratio of the distribution confirms the
excellent sensitivity to these pulses. The observed single electron
event rate was 5.7~s$^{-1}$. The $x,y$ spatial distribution of these
events, shown in Figure~\ref{SEposition}, is relatively uniform within
the selected volume, although a bias is observed at larger radii. This
is attributed to a small and well-understood tilt of the detector of a
few mrad, which affects the thickness of the gas layer (and, to a
lesser degree, the electric field strength), causing a systematic
variation in the detection efficiency at large radii.

\subsection{The WIMP science runs}
The first science run of ZEPLIN--III acquired WIMP-search data for
83~days. Details of the data acquisition and primary dark matter
analyses are described elsewhere
\cite{akimov07,lebedenko09a,lebedenko09b,akimov10a}. Two populations
of small S2-like pulses were observed in the waveforms. One {\em
follows} large S1 or S2 signals and is clearly induced by the VUV
luminescence \cite{edwards08}; this population allows for an accurate
characterisation of single electron clusters. Another can be observed
in the quiet pre-trigger region of the waveforms, {\em before} S1
pulses, and may be unrelated (at least directly) to energy depositions
from interactions in the active volume; a `spontaneous' rate can be
measured for this population.\footnote{This terminology does not
preclude a radiation-induced origin for these events; its use in
quotation marks highlights that no event was detected within the
maximum drift time of the time projection chamber.} We associate these
clusters with those studied in the DSER dataset. We now treat both
types of signal in turn.

\subsubsection{Photon-induced signals and their application}
\label{postS1}

In the FSR, a significant fraction of recorded events contained only
an S1 pulse due to interactions below the cathode grid from $\beta$
and $\gamma$ backgrounds from the PMT array. A reverse electric field
there does not allow electrons to drift to the sensitive volume and
produce S2 signals. These otherwise clean waveforms provide the ideal
dataset to search for the small single electron signals and to
subsequently test the photoionisation production mechanism suggested
previously \cite{edwards08}. The FSR waveforms were processed by the
same data reduction algorithms as those implemented for the dark
matter search data, but with some reduction parameters optimised for
the clustering of even smaller S2-like signals.

Instead of the photoelectron counting method employed with the DSER,
in this analysis the areas of clustered pulses were integrated in each
channel; these were then converted into photoelectron numbers by using
the single photoelectron response for each phototube determined by the
method described in Ref.~\cite{neves11}. The pulse area distribution
shown in Figure~\ref{SEhistos} (right) yields a mean
31.4$\pm$0.07(stat) photoelectrons per electron, which is only
$\approx$10\% higher than the value determined by the counting method
employed with the DSER dataset. The standard deviation is also
slightly larger in this instance (which is expected for a dataset
acquired over a much longer period and which includes uncertainties in
the single photoelectron response determined for each channel), but
neither is far from the Poisson limit of $\approx$5.5
photoelectrons. As had been observed in ZEPLIN--II, the frequency of
these events is proportional to the number of VUV photons in the
preceding S1 pulse, which points to a photoionisation
origin~\cite{edwards08}.

\begin{figure}
  \begin{center}
    \includegraphics[width=3.5in,clip=on]{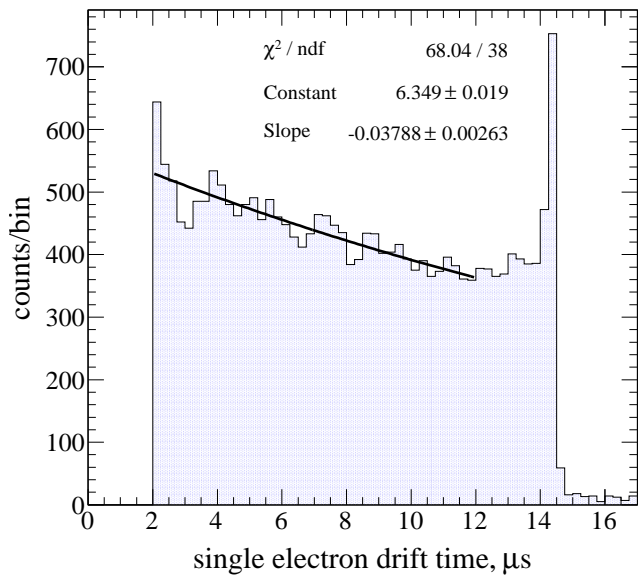}
  \end{center}
  \caption{Drift time (i.e.~depth) distribution of S1-induced single
  electron pulses for one day of FSR data. The cathode grid, located
  at 14.5~$\mu$s drift time, produces an excess due to photo-electric
  emission from the stainless steel wires caused by the scintillation
  VUV photons. The first 2~$\mu$s were excluded since these are
  contaminated by PMT afterpulsing following large S1 pulses. The time
  constant of the exponential fit gives an electron lifetime in the
  liquid $\tau$=26.4$\pm$2.0~$\mu$s.}
  \label{SEdtime}
\end{figure}

The distribution of time elapsed between S1 and the single electron
cluster is shown in Figure~\ref{SEdtime} for a typical day's data. The
spike in event rate near 14.5~$\mu$s is due to photo-electric
production by S1 photons incident on the cathode grid (we find its
probability to be broadly consistent with the quantum efficiency of
metals at VUV wavelengths). The exponential trend observed away from
the cathode is consistent with an electron attachment explanation
based on an independent lifetime measurement (23.2$\pm$1.5~$\mu$s for
that day, obtained as explained below). This suggests that S1-induced
electron emission might enable a good measurement of this parameter,
provided that these electrons are uniformly distributed in the liquid
depth. Both the spatial distribution of energy depositions in the
chamber and the attenuation length for VUV light must be considered in
determining whether the VUV flux from scintillation is constant across
the liquid. The small vertical dimension of the chamber and the
uniform horizontal distribution of this background ensure that this is
indeed the case. The photon mean free path for photoionisation was
estimated in Ref.~\cite{edwards08} as $\sim$km, so a uniform VUV flux
implies a constant probability for electron appearance with depth,
which translates to an exponential survival probability as observed in
Figure~\ref{SEdtime}.

This depth distribution should provide a measurement of the free
electron lifetime in a rather robust and expedite way, relying solely
on the science waveforms themselves (as in this instance) -- rather
than on dedicated calibration datasets. Typically, this calibration is
conducted by fitting the S2/S1 ratio as a function of drift time
(S1-S2 time separation) in response to $^{57}$Co or other external
$\gamma$-ray sources. In this instance it is not straightforward to
prevent (or correct for) saturation of the PMT readout for very large
S2 pulses, which may easily contain upwards of 10$^5$
photoelectrons. The very high VUV photon rates in the chamber may also
affect the operation of these detectors in other, more subtle ways
(e.g.~photocathode charging~\cite{araujo04}). The single electron
signature provides a mechanism which involves only very small pulses
which are anyway present in the data.

\begin{figure}
  \begin{center}
    \includegraphics[width=5in,clip=on]{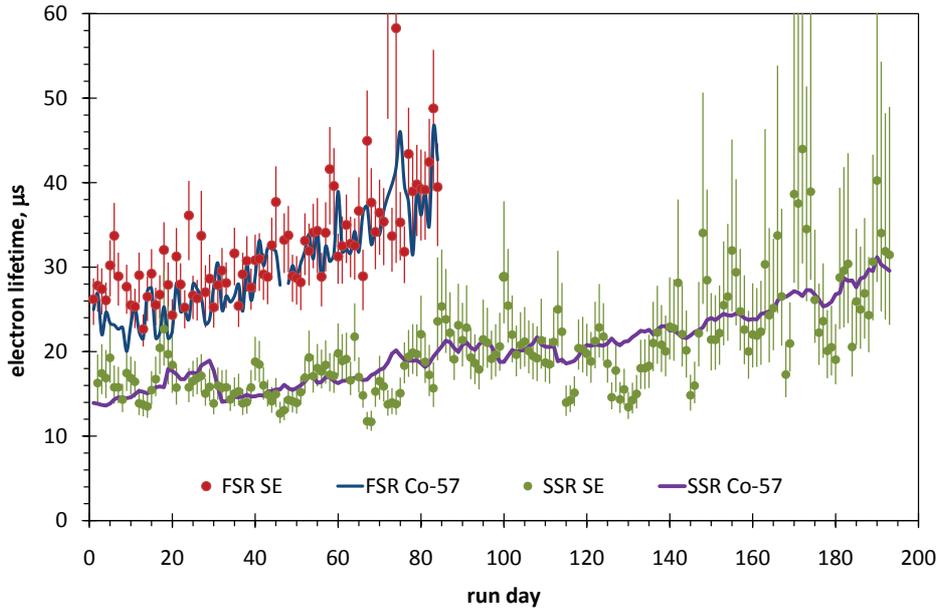}
  \end{center}
  \caption{Historical evolution of the free electron lifetime in the
    liquid xenon measured from single electron signals (markers) and
    from the daily $^{57}$Co calibration in the FSR and the first 200
    days of the SSR (lines). The latter data were corrected
    independently for PMT saturation for large S2 signals but were not
    scaled otherwise.}
  \label{SElifetime}
\end{figure}

Figure~\ref{SElifetime} compares historical lifetime measurements with
this method with those obtained from the daily calibration with
$^{57}$Co for both science runs of ZEPLIN--III. The agreement is very
good in both cases. We anticipate that this new method may be
especially useful for next-generation two-phase xenon and argon
experiments, where routine irradiation of the large WIMP targets with
external sources will be even harder to achieve. Even if not used on
its own, this technique may be useful to assess any biases in the
traditional measurement.

\subsubsection{`Spontaneous' signals in the science runs}

Another single-electron search was conducted in segments of the
WIMP-search waveforms preceding S1 pulses, totalling $\sim$200~s of
live time. These events, like those found in the DSER, have no
traceable correlation with particle interactions in the detector. The
rate for this population (within the 60~mm reconstructed radius)
stabilised at $\sim$20~s$^{-1}$ approximately one month into the FSR
(starting out twice as high initially). This is higher than the rate
of 5.7~s$^{-1}$ measured in the DSER. However, we note that in this
instance the length of waveform available for analysis (16~$\mu$s)
would not allow enforcing a 20~$\mu$s inhibit time from a candidate
pulse.

The experiment was upgraded after the first run, with new PMTs with
considerably lower radioactivity replacing the previous array,
reducing the electron-recoil background of the experiment by a factor
of 18~\cite{araujo11}. The $>$300-day second science run (SSR) started
in mid 2010. In these data, we observed a very significant reduction
in the `spontaneous' component, to a level of
$\sim$1~s$^{-1}$. However, the reduction in background was not the
only difference between the FSR and SSR datasets: the electrode
voltage configurations were also different, producing a slightly lower
electric field in the drift region (13\%) and significantly lower at
the cathode wire surface (see Table~I). Therefore, one cannot conclude
for a background-related explanation for this reduction based solely
on this observation.

A thorough study of the electric field dependence of `spontaneous'
emission was not possible prior to decommissioning of the experiment
at the Boulby mine. Nevertheless, a short dedicated dataset had been
acquired in a similar fashion to the DSER described previously after
the end of the SSR data-taking, but this time with a $^{137}$Cs source
located above the detector; the energy deposition rate in
$\gamma$-rays was some 300 times greater than background. The
resulting `spontaneous' rate was 118~s$^{-1}$. Significantly,
reanalyses of this dataset requiring clear 40~$\mu$s and 60~$\mu$s
periods preceding the single electron cluster (rather than the
standard 20~$\mu$s) yielded lower rates of 55~s$^{-1}$ and
44~s$^{-1}$, respectively. We may conclude, therefore, that not only
does the total `spontaneous' rate depend very significantly on the
rate of energy deposition in the target, but it also decreases with
increasing time delay from those energy depositions.

\subsection{Causes and implications of `spontaneous' electron emission}

This signal will provide a background for applications aiming to
explore the ionisation channel in the few-electron regime, and it is
therefore important to ascertain its (possibly multiple)
causes. Amongst those we single out i) delayed emission due to
lower-than-unity cross-phase extraction probability and ii) field
emission from the cathode grid.

The correlation between the `spontaneous' emission and the
$\gamma$-ray event rate in the SSR and the $^{137}$Cs datasets --
acquired with similar field configurations -- argues for a delayed
emission mechanism, which is further supported by the decrease in rate
for progressively longer inhibit periods. The cross-phase emission
probability \cite{gushchin79} was 83\% under the FSR/DSER conditions
and 66\% in the SSR/$^{137}$Cs runs. Most electrons heated by the
electric field to energies above the field-dependent surface potential
barrier are emitted promptly into the gas phase. Those left on the
surface quickly thermalise with the medium, but the most energetic
ones can still be emitted as thermal electrons, albeit with lower
probability~\cite{bolozdynya99}. Thermal emission is naturally
curtailed by the finite electron lifetime in the liquid, which was
45.4~$\mu$s for the $^{137}$Cs dataset. The observed decrease in
emission rate is broadly consistent with this lifetime. Under this
model, the $\sim$3$\times$ discrepancy between the rates observed in
the FSR and DSER datasets can be attributed to the longer inhibit
period enforced in the latter case.

We also conclude that field emission from the stainless steel cathode
wires (100~$\mu$m diameter, 1~mm pitch) can contribute to the
spontaneous single electron rate to some degree. The field strength at
the surface of the wires is actually comparable to the so-called
`applied breakdown field' which characterises the onset of macroscopic
field emission in metals (see, e.g., \cite{noer82,cox77}). This
breakdown field is determined by electrode irregularities and is some
two orders of magnitude lower than the the critical breakdown field
for the material, which is determined by its work function (although
the latter is, in this instance, decreased by 0.67~eV due to immersion
in the liquid xenon~\cite{gordon94}). The electrode voltage
configuration changed between the first and second science runs;
although the electric field strength in the liquid xenon bulk was only
13\% lower in the SSR, the field near the top of the wires, where
field lines connect with the anode, actually decreased from 60~kV/cm
to 38~kV/cm~\cite{garfield} (this is mainly a consequence of the lower
voltage applied to the cathode, which also establishes a reverse field
region to a second wire grid located underneath it which shields the
PMT input optics). The field dependence of the emission current is
extremely steep ($\propto$$E^2e^{E}$ in the Fowler-Nordheim
model~\cite{fowler28}) and, being a quantum tunnelling effect, no
strict threshold can be defined. For example, the two emission sites
studied in \cite{cox77} (caused by an insulating particle on the metal
surface and by a surface imperfection) would cause far higher rates
than observed here, although no measurable emission would result from
the perfect wires. Therefore the 50\% lowering of the field near the
surface of the wires between runs could easily explain a reduction in
electron emission rates of several orders of magnitude and may
contribute to a baseline rate $\lesssim$1~s$^{-1}$.

Based on these findings, controlling single-electron emission will
require a low background experimental set-up, achieving near-unity
emission probability for hot electrons (i.e.~electric fields
$\gtrsim$5~kV/cm below the surface) and perhaps maintaining a lower electron
lifetime (so as to limit the delayed emission of thermal
electrons). The first (and possibly second) of these requirements will
be challenging for detectors operating in surface laboratories. Field
emission from the cathode is also identified as a viable production
mechanism, but it is perhaps the more benign of the two since careful
electrode design should be able to avert significant spontaneous
rates.

\section{Sensitivity to coherent neutrino-nucleus scattering}

We now assess the feasibility of measuring coherent neutrino-nucleus
scattering using only the low-threshold ionisation channel in a
two-phase xenon detector, with ZEPLIN--III as the working
example. Having characterised the response to single electrons and
identified the foremost sources of background for this signature, we
must consider the uncertain ionisation yield for nuclear recoils in
liquid xenon, which has not been measured experimentally below a few
keV. However, present data (see also, e.g.,~Fig.~5 in
Ref.~\cite{horn11}) suggest that the energy required to extract the
first ionisation electron with appreciable probability is likely to be
$<$1~keV, and maybe significantly less.

For lack of experimental data at very low energies we resort to
extrapolating existing ionisation yields in an {\it ad hoc} manner. We
adopt the yield curve obtained for the ZEPLIN-III FSR~\cite{horn11}
down to 4~keV, which is in good agreement with other data. This yields
a maximum of 6.5~e/keV at that energy, below which it is forced to
inflect and vanish for zero recoil energy along a 3$^{rd}$ order
polynomial. It takes 540~eV to create the first ionisation electron in
this parametrisation. (A more optimistic scenario was also considered
which matched the former above 6~keV but followed instead the last
three data points from Ref.~\cite{manzur10} below that energy; this
curve peaks at $\approx$10~e/keV for 2.7~keV, with the first electron
emitted at 350~eV. However, the conclusions of this study remained
unchanged, and we present only detailed results obtained with the
former scenario.)

\subsection{Coherent neutrino scattering}

Coherent neutrino-nucleus elastic scattering is a predicted high-rate
interaction of the Standard Model whereby neutrinos interact with the
nucleus as a whole through $Z^{0}$ exchange~\cite{freedman74}. This
flavour-blind process measures the total neutrino flux and could have
important practical applications (e.g.~nuclear reactor monitoring) as
well as probe new physics. For example, if this rate were observed to
oscillate this could provide evidence for the existence of sterile
neutrinos. Existing detectors exploit the elastic scattering of
neutrinos off electrons or inelastic scattering from individual
nucleons, in processes which produce much higher energy depositions
(but lower rates) for neutrinos of the same energy. Several studies
have proposed to attempt the extremely challenging detection of
coherent neutrino scattering
(e.g.~\cite{scholberg09,collar09,anderson11}), and notably that in
Ref.~\cite{hagmann04} addresses the type of signature studied here. In
these studies, xenon has been consistently pointed out as a favourable
detection medium when the combined effect of recoil energies and
scattering rates is considered.

The maximum recoil energy, $E_{r}^{m\!a\!x}$, from a neutrino with
energy $E_{\nu}$ is inversely proportional to the mass of the target
nucleus ($M$):
\begin{equation} \label{eq:erec}
  E_{r}^{m\!a\!x}=\frac{2E_{\nu}^{2}}{M+2E_{\nu}} \,.
\end{equation}
The differential cross section is given by:
\begin{equation} 
  \label{eq:cross}
  \frac{d\sigma}{dE_{r}}=\frac{G_{F}^{2}}{4\pi}Q_{W}^{2}M
  \left(1-\frac{ME_{r}}{2E_{\nu}^{2}}\right)F^{2}\left(Q^2\right)\,,
\end{equation}
where $G_{F}$ is the Fermi constant, $F(Q^2)$ is the form factor at
four-momentum $Q$ and $Q_{W}$=$N\!-\!(1\!-\!4\sin^2(\theta_W))Z \sim N$ is the
weak charge for a nucleus with $N$ neutrons and $Z$ protons, with
$\theta_W$ the weak mixing angle; $Q_W^2$ enhances the scattering rate
through coherence and favours heavier elements.

\subsection{Signal rate estimation}

\begin{figure}[ht]
  \begin{center}
    \includegraphics[width=3.5in]{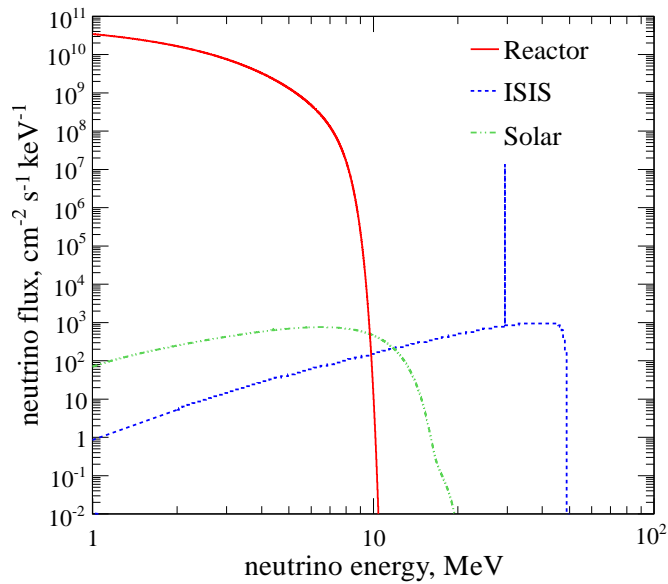}
  \end{center}
  \caption{Neutrino fluxes considered in this work; the solar flux
	includes $^{8}$B and `hep' contributions~\cite{bahcall95}; the
	reactor flux from Ref.~\cite{kopeikin97} is adopted; the ISIS
	spectrum includes the beam-prompt $\nu_{\mu}$ line at 29.8~MeV
	and the two continuum contributions from $\bar{\nu}_{\mu}$ and
	$\nu_{e}$~\cite{armbruster98} (which follow a muon decay
	timescale with $\tau$=2.2~$\mu$s).}
  \label{flux}
\end{figure}

Assuming a recoil energy threshold $E_{th}$=0.5~keV,
Eq.(\ref{eq:erec}) places a lower bound of 5.5~MeV on the detectable
neutrino energy for a xenon target. On the other hand, maintaining
scattering coherence requires small momentum transfer ($qR\!<\!1$,
where $R$ is the nuclear radius and $q$ is the three-momentum),
determining an upper bound of $\sim$50~MeV for the neutrino
energy. This range limits the neutrino sources that can be detected in
this way. Seeking those with highest flux, we consider the following
-- represented in Figure~\ref{flux}:
\begin{itemize}
\item Solar neutrinos: a fraction of the solar neutrino spectrum,
  namely the $^8$B and `hep' ($\rm{^3He}+p \to \rm{^4He} + e^+ +
  \nu_e$) contributions, delivers high flux in the interesting energy
  range. We assume the Bahcall model for the neutrino
  spectra~\cite{bahcall95}. (This case is included mainly for
  reference, since the scattering rates are very small indeed.)
\item Reactor antineutrinos: we address the possibility of placing a
  detector like ZEPLIN--III some 10~m away from a 3~GW nuclear
  reactor, where the neutrino flux would be
  $\sim\!4\!\times\!10^{13}$~cm$^{-2}$s$^{-1}$; we adopt spectra from
  Refs.~\cite{kopeikin97} and \cite{vogel89}. Several experiments have
  operated at similar distances from a nuclear reactor core and new
  ones are being considered \cite{avenier89,porta10,akimov09}.
\item Stopped pion sources are promising for these
  studies~\cite{scholberg06}. We consider neutrinos produced at the
  800~MeV proton beam at the ISIS facility (Rutherford Appleton
  Laboratory), with a pulse repetition rate of 50~s$^{-1}$ and beam
  current of 200~$\mu$A~\cite{armbruster98}. The neutrino flux 10~m
  away from the target is estimated as
  $\sim$1$\times$10$^7$~cm$^{-2}$s$^{-1}$ per flavour.
\end{itemize}

\begin{figure}[ht]
  \begin{center}
    \includegraphics[width=3.5in]{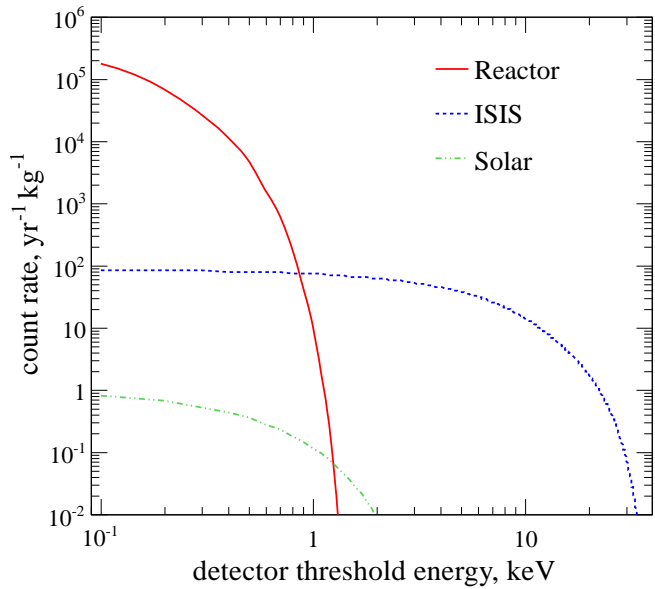}
  \end{center}
  \caption{Expected integral rate of coherent neutrino-nucleus
    scatters in liquid xenon as a function of detector threshold for
    nuclear recoils.}
  \label{evt_rate}
\end{figure}

The integral event rate is given by:
\begin{equation} 
  R(E_{th})=N_{t}\int_{0}^{\infty}dE_{\nu}\Phi\left(E_{\nu}\right)
  \int_{E_{t\!h}}^{E_{r}^{m\!a\!x}}dE_{r}
  \frac{d\sigma\left(E_{\nu},E_{r}\right)}{dE_{r}}\,,
  \label{eq:rate}
\end{equation}
where $N_{t}$ is the number of target atoms,
$\Phi\left(E_{\nu}\right)$ is the neutrino flux and $d\sigma/dE_{r}$
is the differential cross-section given by
Eq.(\ref{eq:cross}). Figure~\ref{evt_rate} shows the computed event
rates of coherent neutrino-nucleus scatters in xenon for the neutrino
fluxes mentioned above.

From these event rates the neutrino signal induced in the detector is
evaluated. We consider 10~kg$\cdot$year as a reasonable dataset
(1.5-year run with the nominal ZEPLIN--III fiducial mass of
6.5~kg). For each neutrino-nucleus interaction, very few (if any) S1
VUV photons and some ionisation electrons would be generated, but the
S1 signal is most likely lost. However, due to the sensitivity
achieved in the ionisation channel, a measurement can be made even if
only one electron is extracted from the interaction point. For S2-only
events the depth information normally afforded by the time projection
chamber is lost (but $x,y$ positional information is still available)
and so is particle discrimination by S2/S1 ratio. For this reason, it
is essential to record S1 pulses whenever present; crucially, this
will also help reject important backgrounds (e.g. radioactive
contamination on the cathode grid).

\subsection{Backgrounds}

We assume that external radioactivity neutrons and those associated
with the reactor or beam are suitably mitigated with sufficient
hydrocarbon shielding combined with an anti-coincidence system
surrounding the xenon target -- like that deployed in the second run
of ZEPLIN--III~\cite{akimov10b,ghag11}. A veto detector will also tag
internal neutrons efficiently and reduce the effect of cosmogenic
neutrons to manageable levels.

A dominant background will arise from the single electron emission
mechanisms discussed previously. We consider a rate
$f$$\sim$10~s$^{-1}$, which is comparable, per unit volume, to that
observed (underground) for the SSR. Hopefully, a higher extraction
field and a longer inhibit period to preceding events can mitigate the
potentially higher event rate expected with a surface deployment. We
include the probability of $n$-electron coincidences according to
$nf^{n}\Delta t^{n-1}$, where $\Delta t$=1~$\mu$s is the typical
single electron signal duration. We point out that the coincidences
rate could be lowered significantly ($\sim$10 for a detector like
ZEPLIN--III) with more sophisticated position algorithms.

A more challenging background arises from low-energy electron recoils
due to $\beta$ and $\gamma$ radiation (internal and cosmogenic,
assuming substantial $\gamma$-ray shielding). The SSR background rate
of $\sim$1~event/kg/day/keV, measured and modelled in
Ref.~\cite{araujo11}, is considered. The average energy per emitted
electron from an electron recoil is $\sim$50~eV -- obtained from the
W-value for LXe at zero field~\cite{takahashi75} and relevant
field-dependent extraction (from track) and surface emission
probabilities. This background represents 180~events/electron in a
10~kg$\cdot$year dataset. Above $\sim$2~keV in S1 (corresponding to
$\sim$50-electron S2 signals from electron recoils) efficient
discrimination by S2/S1 ratio will become possible.

This background can be suppressed significantly in an ISIS experiment
by exploiting the pulsed nature of the beam: a triggered measurement
lasting for one drift length ($\sim$20~$\mu$s) per proton bunch (20~ms
period) would effectively reduce most non beam-related backgrounds by
a factor of $\sim$1,000. In addition, the depth coordinate can still
be recovered with reasonable accuracy by replacing the S1 pulse by the
trigger signal, enabling fiducialisation and self-shielding of
external backgrounds -- this will work well for the beam-prompt
neutrinos; the continuum components will be delayed by a few $\mu$s,
reducing the timing resolution of the chamber.  For a reactor
experiment, `on/off' background subtraction will be far less
effective.

\subsection{Predicted observable spectra}

\begin{figure}[ht]
  \begin{center}
    \includegraphics[width=4.5in]{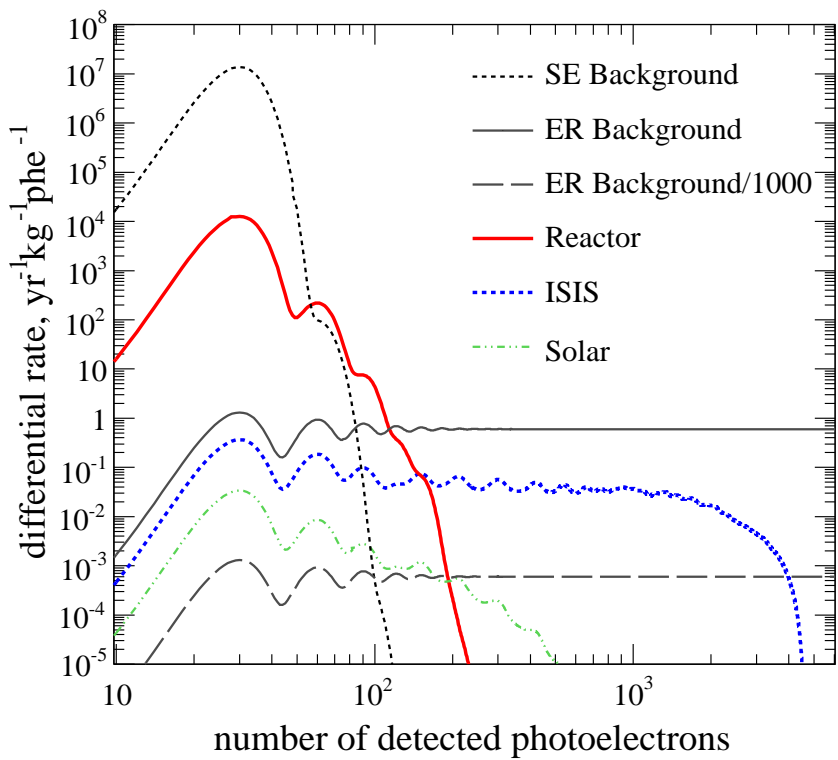}
  \end{center}
  \caption{Ionisation spectra expected from coherent neutrino
    scattering in ZEPLIN--III exposed to different neutrino
    sources. Single electron and electron recoil backgrounds are also
    shown. The peak structure reflects discrete numbers of ionisation
    electrons measured by electroluminescence.}
    \label{cnsrates}
\end{figure}

In Figure~\ref{cnsrates} we present photoelectron spectra predicted
for neutrino interactions and for the two dominant backgrounds, as
would be observed in ZEPLIN--III. Individual peaks represent
ionisation electrons detected by electroluminescence; we assume a
yield of 30~photoelectrons per electron and Poisson variance.

As the figure suggests, the neutrino signal must be searched above
$\gtrsim$3~electrons due to the single electron background -- although
this will be improved with multiple-cluster resolution in $x,y$ using
advanced position algorithms. The electron recoil background becomes
significant above that threshold, but the reactor signal is clearly
salient near 100~phe. Unfortunately, its spectrum does not extend to
1,500~photoelectrons in the S2 channel (50 electrons), which would be
required for a detectable S1 pulse from an electron recoil thus
enabling discrimination by S2/S1 ratio.

For a reactor experiment, the number of events expected in a
10~kg$\cdot$yr dataset above a 75~phe threshold is of order 3,000
(1,000 above 90~phe). The electroninc background is $\sim$200 events
over the relevant range. These values are sensitive to the shape of
the antineutrino reactor spectrum and the ionisation yield for low
energy recoils. The number of signal events changes only by a few tens
of percent when the antineutrino spectrum of Ref.~\cite{vogel89} is
used instead. However, the optimistic ionisation yield scenario
mentioned previously would increase the signal rate ten-fold. This is
therefore an appropriate level of systematic uncertainty to frame
these calculations as far as the signal is concerned. The background
in the reactor environment will depend critically on the shielding
efficiency, which must be higher than typically required for
underground WIMP searches. We note also that self-shielding will not
be very effective in the absence of the depth coordinate. In spite of
this, the reactor experiment seems viable.

Signal rates at the stopped pion source are lower but the spectrum
does extend to considerable energies. Significantly, in this instance
the electron recoil background can be reduced by three orders of
magnitude with a beam-coincident measurement. Discrimination should be
possible for approximately half of the events, when S1 pulses are
expected. 3D position reconstruction should be achieved to a
resolution of a few mm in the depth coordinate by using the trigger
signal to define zero drift time. The integrals above 75~phe are
$\sim$700 and $\sim$10 background events over the same energy range
for a 10~kg$\cdot$year exposure. This result is not very sensitive to
the ionisation yield at low energy. In conclusion, the spallation
source produces very encouraging numbers.

\section{Conclusions}

In this article we report studies of single electron emission using
data from the ZEPLIN--III dark matter experiment and assess two
applications which exploit the ability of xenon emission detectors to
sense the quantum of response in ionisation. Electroluminescence
signals due to single electrons emitted from the liquid xenon target
were analysed. We showed that such pulses, containing an average of
$\sim$30~photoelectrons in the FSR configuration, can be detected with
very high signal-to-noise ratio and exhibit near-Poisson variance.

The source of single electrons following the scintillation generated
by sizable energy depositions in the liquid xenon is thought to be
photoionisation of impurities by the VUV photons. We demonstrated that
this signature can be used to obtain a very robust measurement of the
free electron lifetime in the liquid phase from science data
themselves.

In addition to this photon-induced population, `spontaneous' electron
emission was also studied and attributed to delayed emission of
thermal electrons trapped at the surface. Clearly, the physics of the
emission process at the liquid-gas interface is a topic deserving
further study. It was also concluded that field emission from cathode
wire grids could be significant and should be considered when
designing these detectors. These single electron pulses determine a
$\sim$3 electron threshold for experiments exploiting the ionisation
channel below the scintillation threshold.

We assessed the feasibility of two such experiments related to the
detection of coherent neutrino-nucleus elastic scattering, a Standard
Model process not yet observed, using realistic signal characteristics
and backgrounds. We found that the signal from the nuclear reactor
scenario considered in this study is salient above the electron recoil
background, with $\gtrsim$2,000 events/(10~kg$\cdot$yr) expected above
a $\sim$3-electron threshold. In this instance the signal rate is very
sensitive to the nuclear recoil ionisation yield assumed at low
energies.

The prospects of detecting this elusive neutrino signature with a
beam-coincident measurement at a stopped pion source such as ISIS are
also very encouraging: owing to a harder energy spectrum, a sizable
fraction of events should benefit from electron-nuclear recoil
discrimination; three-dimensional position reconstruction should be
possible; backgrounds would be significantly lower. $\gtrsim$700
events/(10~kg$\cdot$yr) are expected above $\sim$3 emitted
electrons. Even higher neutrino fluxes are likely to be reached in the
coming years: SNS at Oak Ridge~\cite{efremenko09} is expected to
achieve a flux several times higher than that considered here, and
ISIS itself may have its capability improved in a forthcoming
upgrade~\cite{thomason08}.

\acknowledgments 
The UK groups acknowledge the support of the Science \& Technology
Facilities Council (STFC) for the ZEPLIN--III project and for
maintenance and operation of the Boulby underground laboratory.
LIP-Coimbra acknowledges support from Funda\c{c}\~{a}o para
a Ci\^encia e a Tecnologia (FCT) through project grant
CERN/FP/116374/2010 and postdoctoral grants SFRH/BPD/27054/2006,
SFRH/BPD/47320/2008 and SFRH/BPD/63096/\-2009. The ITEP group
acknowledges support from the Russian Foundation of Basic Research
(grant 08-02-91851 KO\_a) and Rosatom (H.4e.45.90.11.1059 from
10-03-2011). We are also grateful for support provided jointly to ITEP
and Imperial from the UK Royal Society. ZEPLIN--III is hosted by
Cleveland Potash Ltd (CPL) at the Boulby Mine and we thank CPL
management and staff for their long-standing support. We also express
our gratitude to the Boulby facility staff for their dedication. The
University of Edinburgh is a charitable body registered in Scotland
(SC005336).

\bibliographystyle{JHEP}
\bibliography{HAraujo}

\end{document}